\documentclass[12pt]{article}
\usepackage[margin=1in]{geometry}                 
\usepackage{setspace}
\usepackage{graphicx}
\usepackage{amsmath}
\usepackage{amssymb}
\usepackage{bm}
\usepackage{url}
\usepackage{epstopdf}
\usepackage{times}
\usepackage{courier}

\usepackage{algorithmic}
\usepackage{textcomp}
\usepackage{xcolor}
\usepackage{subcaption}
\usepackage{diagbox}
\usepackage{tabulary}
\usepackage{mathtools}
\usepackage{listings, multicol}
\title{SAFE: Secure Aggregation with Failover and Encryption}
\author{Thomas Sandholm \and Sayandev Mukherjee \and Bernardo A.~Huberman\\ 
Next-Gen Systems, CableLabs}

\begin{document}
\maketitle

\begin{abstract}
We propose and experimentally
evaluate a novel secure aggregation
algorithm targeted at cross-organizational
federated learning applications with
a fixed set of participating learners. 
Our solution organizes learners in a chain
and encrypts all traffic to reduce the
controller of the aggregation to a mere
message broker.
We show that our algorithm scales better
and is less resource demanding than
existing solutions, while being
easy to implement on constrained
platforms. 

With 36 nodes
our method
outperforms state-of-the-art 
secure aggregation by 70x, and 56x
with and without failover, respectively.
\end{abstract}

\section{Introduction}\label{sec:introduction}
Federated Learning~\cite{kairouz2021} for machine learning (ML) offers an attractive alternative to centralized collection of training data, which is facing increased resistance from regulators, lawmakers, and privacy-conscious consumers~\cite{bonawitz2016, huberman2021}. A typical Federated Learning controller collects and averages model parameters (or updates thereof) from 
local learners. Although this poses less of a privacy loss concern than collecting the raw data,
there is a risk of an attacker reverse engineering the input data by knowing the model
and intercepting the model parameters (or even just updates to the model parameters). The attack could be mounted from an untrusted 
controller or from a compromised controller with the attacker able to intercept all 
messages to and from the controller. 

A number of secure aggregation mechanisms have been
proposed to address this issue in order to guarantee that the controller can only decrypt the average 
or sum of all model parameters across all learners, and not the parameter values at any of the individual learners (e.g. [1]). 

The protocol proposed in~\cite{bonawitz2017} and variations 
thereof~\cite{so2021, choi2020} target learners connected to controllers by unreliable wireless links, but have the following shortcomings:
\begin{enumerate}
\item{Learners may frequently drop out of the protocol at any time because of loss of connectivity, even in the middle of the steps required to complete the computation of the average, and thus failure recovery overhead is incurred even if all nodes complete successfully, e.g. exchanging secrets as well as weights within each aggregation cycle across all learners
is necessary for failover, but makes the algorithm resource inefficient with larger number of number of learners.}
\item{The controller needs to actively participate in the aggregation which exposes the learners 
more to an untrusted or rogue coordinator and also limits concurrency and scalability.}
\item{In case of a failure a large number of nodes need to be contacted to recover, which makes failure cases and privacy loss scenarios complex.}
\end{enumerate}

In contrast, we consider a scenario where 
learners are leaving or entering the system infrequently, and we thus 
provide a simple protocol for a stable system, and a 
couple of mitigating algorithms that kick in when there are rare failure. 
We chain nodes in a cycle and thus limit the number of nodes that need to be involved
to mitigate a failure. We also limit the controller involvement to message passing
and progress monitoring.  Another simplification of the operational conditions arises from our assumption that the links between the learners themselves, and the learners and controller, are relatively stable and not subject to drastic variations in link quality.  In practice, this holds if, as we assume here, the learners are either stationary or moving very slowly, so that no links are subject to fading arising from mobility. 

Scaling to a huge number of learners is not a primary concern as simpler models like noise 
injecting differential privacy may be used then.  Furthermore,
cross-organizational (vertical federated learning~\cite{yang2019}) setups are likely to have a fewer participants than, say, an IoT fleet or sensor network in a single organization. The nature of aggregation
also lends itself well to hierarchical federation without loss of privacy, but with more
robustness and scalability.

Finally, we focus on learners that are in the form of devices with constrained computing capabilities (like CMs, WiFi gateways, and IoT devices), unlike the smartphones considered by~\cite{bonawitz2016}.  We study both devices that can run desktop-class software (but are compute-constrained) and an especially constrained class of devices that cannot even run a full machine learning library like sklearn.  In our evaluation we experimentally measure the scalability in nodes and features
on two platforms: an edge compute learner (implemented with Python), and a 
constrained deep-edge learner (implemented with Linux busybox coreutils). 

\section{Related Work}\label{sec:relatedwork}
A secure aggregation protocol for Federated Learning was first outlined in~\cite{bonawitz2016} with the full details published in~\cite{bonawitz2017} (referred to as the BON protocol below), and
then extended in~\cite{choi2020} and~\cite{so2021}.

As stated above, in our primary use case the aggregation nodes are constrained devices
such as IoT or network equipment, e.g. routers, sensors, and access points, and thus they are more stable between rounds and less likely to fail than mobile devices.  In this setting, even though communication is more stable, the compute power on each node may be limited, as these are typically
low-cost, single-purpose devices.

The secure aggregation protocol in~\cite{bonawitz2017} is proposed
for mobile devices, with the assumptions that communication
is expensive, communication link failures are common, aggregation vectors
are high-dimensional, and the server is not trusted. Like our approach, the
server (controller) routes messages between participants,
but unlike our approach, it also computes the final result.

The main difference between our approach and that of~\cite{bonawitz2017} is the way the aggregate is masked.  In particular, we only use a random mask from the initiator node in the chain (see Sec.~\ref{sec:basic_protocol_no_failures}), 
whereas~\cite{bonawitz2017} employ a pairwise mask between each pair of
nodes. Hence, if a non-initiator fails we can recover more
efficiently without having to worry about unmasking, and without having
to contact all remaining nodes to recompute secrets. The drawback in our approach is
that if the initiator fails, we need to rerun the protocol from the beginning,
but based on our assumptions this should be a rare scenario.

The masking is made even more complicated in~\cite{bonawitz2017} in that
a delayed response from a node can be easily unmasked by asking all other nodes
for their masks with the presumed failing node. So the nodes have additional
masks to ensure they do not reveal the local parameters to the server in case of a false
failure.  On the other hand, our proposal does not have this complication since we only have a single mask maintained by the initiator and never revealed to the server or any other participating node.

On a high level, the BON protocol requires four rounds of communication (public key sharing,
weight sharing, masked secret sharing, and unmasking with failure secret recovery). 
In our protocol (SAFE) we have the same first
two rounds (public key sharing, weight sharing), but the masked secret sharing is not needed
and the failure recovery only involves a single node (retransmitter) unless the
initiator failed, which would involve rerunning the protocol~\footnote{The definition of a round
here differs slightly from the presentation in~\cite{bonawitz2017} and is more in line with
the practical implementation we use from \url{https://github.com/ammartahir24/SecureAggregation/}}.

Finally, we also note that~\cite{bonawitz2017} uses a real implementation of the protocol in simulations
but not a full distributed system to implement the communication, and hence their measured
scalability numbers in terms of number of nodes look quite different from our evaluations
that uses a multi-threaded client or separate clients on Wi-Fi access points
and a server over a REST/HTTPS protocol. Our setting is kept the same across all benchmarks
measured, but the scalability numbers between our evaluations and 
theirs cannot be directly compared, for this reason.

In~\cite{choi2020} the authors also recognize the scalability and computational 
complexity of the BON protocol and improve on it by splitting
up the aggregation across subgraphs. Such subgraph splitting could also be applied on top
of our protocol, and our subgroup feature does exactly that. Our experiments have shown
that the BON primitives scale poorly even with a very limited set of nodes, and thus
it is more efficient to re-design the core protocol in our case.

In~\cite{so2021} the authors also propose a topology overlay across the BON primitives, called
Turbo-Aggregate,
and in particular a circular subgroup one as opposed to a sparse subgraph like in~\cite{choi2020}.
Although our approach also relies on a circular topology and subgrouping, our masking primitives
are distinctly different from the ones in BON not to suffer from the same scalablity issues
or computational complexity (quadratic in number of users), and thus subgrouping is only
required in our case when there is a large number of users.

Turbo-Aggregate applies additive secrets coordinated 
by the server to each user and subgroup as opposed to
being local to the initiator as in our case. Hence, our model puts less of a burden and trust
liability on the server or coordinator, and is easier to recover from in case of failures.
The implementation complexity of our solution is also simplified due to the fact that no
k-of-n secret sharing is necessary. The parallel computation in subgroups is similar
between the Turbo-Aggregate approach and ours. It could in theory be possible to apply the Turbo-Aggregate
topology on top of our SAFE algorithm, as opposed to the BON primitives.

Another protocol called Fast Secure Aggregation~\cite{kadhe2020} addresses the susceptibility to attack by non-adaptive adversaries of both Turbo-Aggregate~\cite{so2021} and a modification~\cite{bell2020} of the original BON protocol that replaces the complete computation graph of~\cite{bonawitz2017} by a sparse random graph.  However, Fast Secure Aggregation has the same communication complexity as BON without the strong privacy guarantees of BON, being designed for the honest-but-curious setting instead of Byzantine clients.

\section{Motivation for the Proposed Protocol}\label{sec:model}
Our work draws inspiration from the large body of work on Secure Multiparty Computation (SMC)~\cite{evans2017}. Although SMC methods ``help protect intermediate steps of the computation when multiple parties perform collaborative machine learning on their proprietary inputs, ... SMC techniques impose non-trivial performance overheads and their application to privacy-preserving deep learning remains an open problem''~\cite{shokri2015}.

The cryptographic-based protocol of~\cite{bonawitz2017} was a step forward in SMC applied to deep learning in that it represents the first secure privacy-preserving protocol for federated training of deep learning models. However, it involves a complex set of interactions and cryptographic functions, typically not available or too slow to run on the constrained devices of interest in the present work, e.g. network equipment like routers and cable modems (CMs).  This cryptographic complexity exists to combat the presumed complete untrustworthiness and maliciousness of the learners.

On the other hand, the aggregation protocol in the present work was inspired by a discussion of algorithms and assumptions regarding trustworthiness in~\cite{huberman2021}. In contrast to~\cite{bonawitz2017}, in~\cite{huberman2021} and also in the present work, the operational scenario assumes that the learners have a certain degree of trustworthiness in and amongst themselves, and are not anonymous public devices (like smartphones owned by consumers, in the case of~\cite{bonawitz2016}) that are outside the control of the operator wishing to run the learning algorithm on them.  For example, the learners could be CMs in a home (which have tamper-protection installed by the cable operator), or IoT devices on a corporate network (therefore authenticated and certified by the administrators of that network), or WiFi gateways with some protection against modification of their firmware.  We also conceive of ``cross-organizational'' federated learning where the learners are spread over multiple enterprises.  In these situations, we expect the enterprises to want to interact with each other productively over time, so each enterprise will be vigilant in enforcing the integrity of its own training data and there will be mutually enforced penalties on any enterprise that seeks to acquire private data from a peer.

The present work adds a distributed systems design and implementation including failover and encryption,
for secure communication across the aggregation chain via a controller,
to the baseline algorithm described in~\cite{huberman2021}. We also present performance results from two implementations targeted at edge nodes and constrained devices.

%Although we will not dwell on the subject, we mention here that all the cited literature as well as our own proposal assume ``classical'' cryptography based on the use of private keys to protect data in transit between the learners, or between the learners and the controller during the operation of the secure aggregation protocol.  However, recent advances in quantum computation permit such conventional cryptographic schemes to be broken.  As outlined in~\cite{huberman2021} and described in detail in~\cite{huberman2020}, data in transit can be secured using quantum protocols for provably (as opposed to algorithmically) secure key distribution via quantum channels.

\section{Method Overview}
\begin{figure}[htbp]
        \centerline{\includegraphics[scale=0.5]{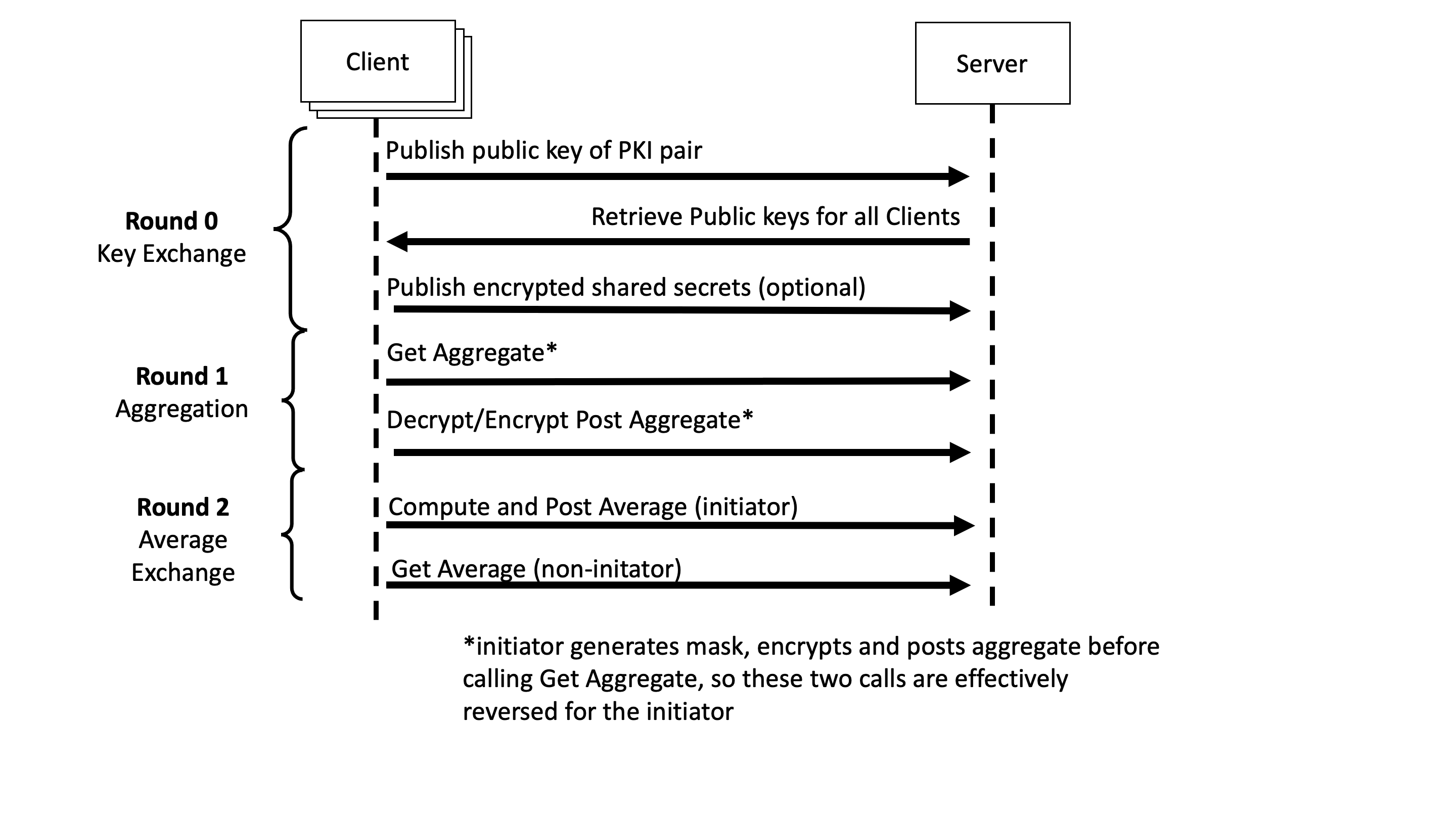}}
        \caption{Overview of our secure aggregation method.}
\label{secaggoverview}
\end{figure}

Our method can be described on a high level as the set of interactions between clients and a server shown 
in Figure~\ref{secaggoverview}.

The first round (Round 0: Key Exchange) is similar to other protocols such as \cite{bonawitz2016} and \cite{kadhe2020}, 
and only involves standard PKI and symmetric key encryption primitives. This round does not need 
to be executed each time a new secure
aggregate is computed, only when new nodes enter the system, which we, as previously mentioned, assume
is a rare occurance.  Shared secrets can be exchanged in this step as an optimization but they can also be shared
as part of the aggregation in the following step.

\begin{figure}[htbp]
        \centerline{\includegraphics[scale=0.5]{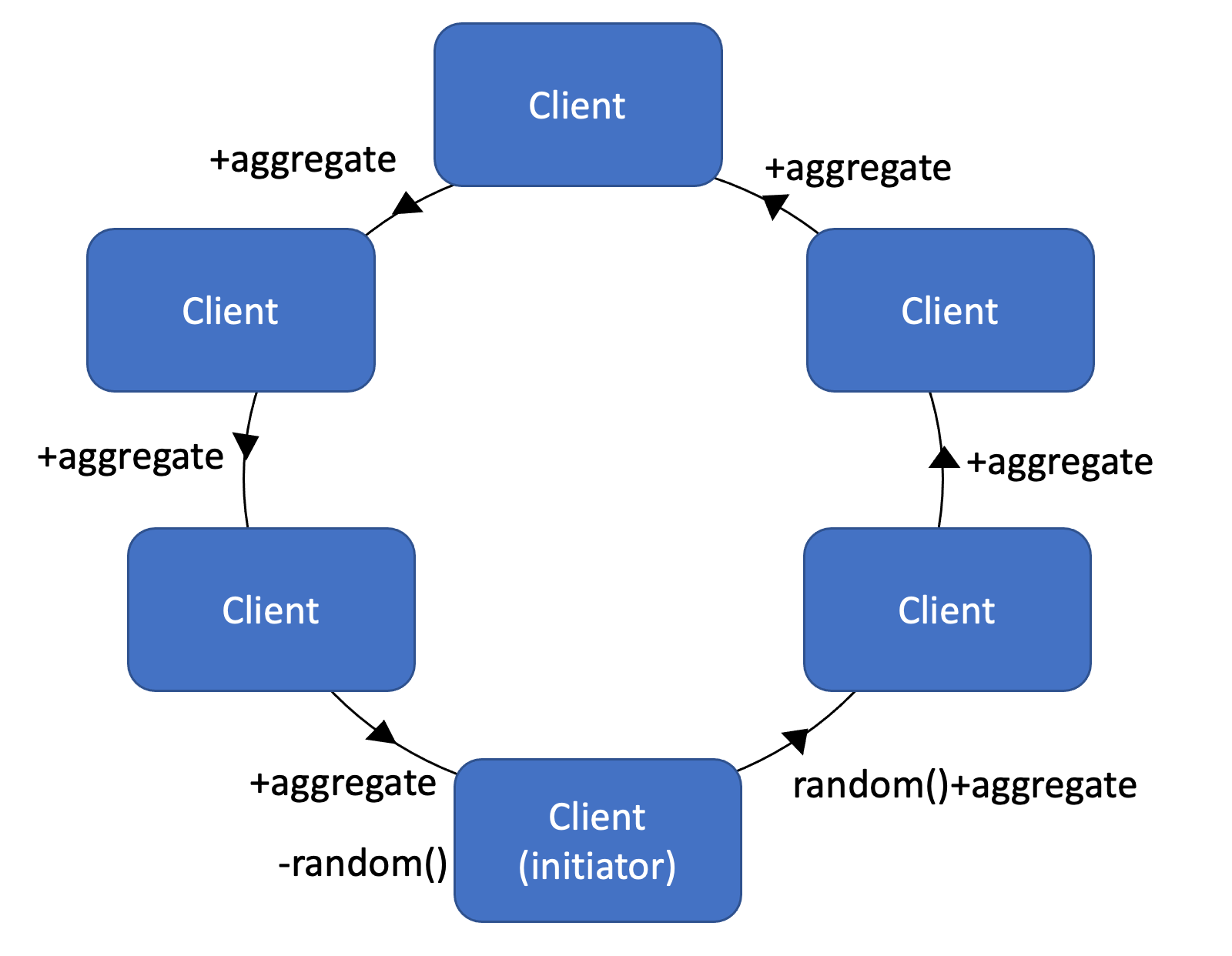}}
        \caption{Round 1 Aggregation: logical communication chain.}
\label{safecirc}
\end{figure}

The gist of our secure aggregation method is encoded in Round 1. Although communication goes through a server to simplify
connectivity between clients, logically a client only communicates with other clients that are 
adjacent to it in an ordered circular chain (see Figure~\ref{safecirc}). The scalability and performance improvements of our method
mainly stem from this round only requiring $O(n)$ messages to be exchaged, $n$ being number of clients, 
as opposed to $O(n^2)$ in other methods. To ensure that the server or other clients cannot eavesdrop
messages all the communication is encrypted with the public key of the receiver.

The computational complexity of our proposed scheme is dominated by the encryption and decryption in Round 1.
If, for example, RSA public key cryptography is used, then the computational complexity of encryption is $O(k^2)$ and 
that of decryption is $O(k^3)$, where $k$ is the number of bits in the modulus of the public key\footnote{\url{https://web.archive.org/web/20071112103441/http://www.rsa.com/rsalabs/node.asp?id=2215}}.

\section{Protocol Design}\label{sec:design}
In this section, we describe the system design of our basic method
and various extended features.

\subsection{Introduction}
Our basic algorithm involves a {\it controller} (a.k.a server) and a set of
at least 3 {\it learners} (a.k.a nodes or clients). One of the nodes
is designated as an {\it initiator} and all other nodes
as {\it non-initiators}. Which node is the
initiator may change between aggregation rounds.
Depending on whether a node is designated as initiator
or not it performs different steps as part of the aggregation.
All nodes have a unique id $[1,2,3..n]$, which denotes their
place in the aggregation chain of length $n$.

\subsubsection{Initiator} 
\begin{enumerate}
\item{The initiator starts off an aggregation
by first generating a large random number that is added
to its local feature vector. Then the resulting
feature vector is encrypted with the public key of
the next node in the chain, and the encrypted aggregate
is posted to the controller.}
\item{Next the initiator waits for the controller notification
that the next node in the chain has consumed the posted 
aggregate}
\item{Then, the initiator waits for the controller notification
that an aggregate is available from the final node in the chain}
\item{Finally the initiator decrypts the final aggregation value
from the last node in the chain, subtracts the random number 
and divides by the number of nodes posting their aggregate values
and post the clear-text average to the controller, for any node to pull}
\end{enumerate}

\subsubsection{Non-Initiator} 
\begin{enumerate} 
\item{A non-initiator starts the aggregation by waiting for
a controller notification that an aggregate is available
from the previous node in the chain.}
\item{When an aggregate is obtained it is decrypted, the
local feature vector is added, and it is encrypted for the
next node in the chain and posted to the controller}
\item{Next the node  waits for the controller notification
that the next node in the chain has consumed the posted 
aggregate}
\item{Finally the node waits for the controller notification
that the average is available and pulls it}
\end{enumerate} 

\subsubsection{Controller} 
\begin{enumerate} 
\item{The controller is responsible for storing messages
sent to target nodes until they are retrieved.}
\item{It also makes sure progress is made, and if there
is a timeout, because a node fails to do its part, the controller
requests the sending node to re-encrypt and resend to
a new target.}
\item{The controller manages distribution of
the computed average across all nodes. If subgroups are used
it also calculates the average across groups.}
\item{If the initiator fails, the controller is responsible
for picking a new initiator.}
\end{enumerate} 

Notifications can be implemented as long-polling or with 
a pubsub service (see Section~\ref{sec:pubsub}).

The controller makes the following operations available to the nodes:
\begin{itemize}
\item{{\bf post\_aggregate(from, to, aggregate)} Node {\it from} sends {\it aggregate}
to node {\it to}.}
\item{{\bf check\_aggregate(node):[empty, consumed, repost]} Nodes call this
operation to check whether a posting has successfully been consumed or a repost
is needed}
\item{{\bf get\_aggregate(node):aggregate} Used by nodes to retrieve {\it aggregate}
sent to them.}
\item{{\bf post\_average(average)} Used by initiator to distribute final result, {\it average}.}
\item{{\bf get\_average():average} Used by nodes to retrieve final result.}
\item{{\bf should\_initiate(node):[true, false]} Used by node after initiator
failure to determine if they should resume the role of initiator in the next attempt.}
\end{itemize}

Given these primitives, we now describe how rounds 1 and 2 in Figure~\ref{secaggoverview} are implemented.

\subsection{Basic Protocol without Failover Procedures}
\label{sec:basic_protocol_no_failures}
\begin{figure}[htbp]
        \centerline{\includegraphics[scale=0.3]{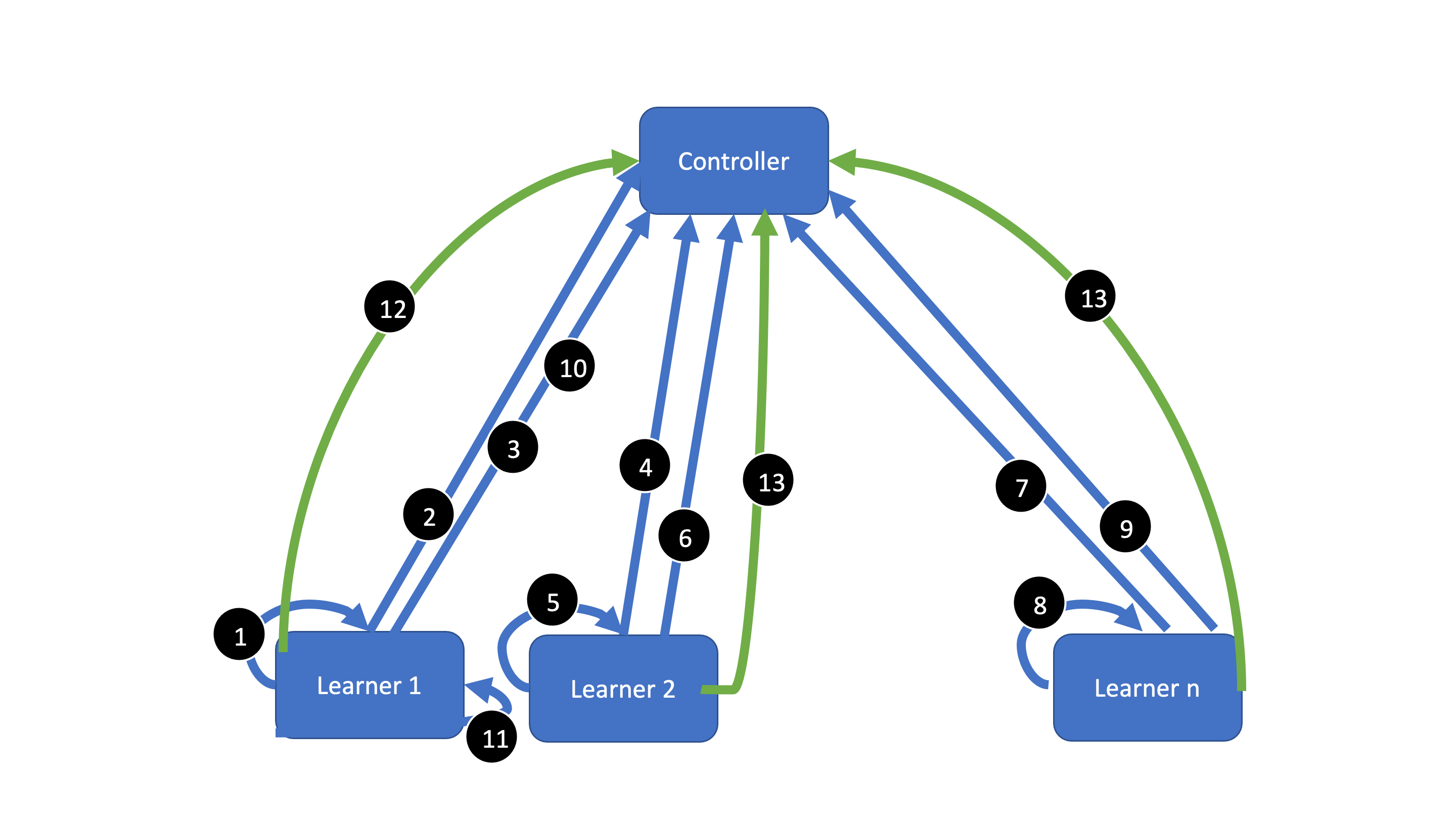}}
        \caption{Basic Algorithm Interactions.}
\label{basicarch}
\end{figure}

The basic interactions between the initiatior, non-initiators and the controller can be seen
in Figure~\ref{basicarch}.
\begin{enumerate} 
\item Learner 1 generates a large random number $R$ and adds its model parameter to it, then encrypts the aggregate ($agg_1$) with the public key of Learner 2.
\item{Learner 1 posts the encrypted aggregate
$enc_2\langle agg_1\rangle$ to the controller by calling \\
post\_aggregate(1, 2, $enc_2\langle agg_1\rangle$)}
\item{Learner 1 starts polling for results from the final learner with get\_aggregate(1)}
\item{Learner 2 calls get\_aggregate(2) and receives \\
$enc_2\langle agg_1\rangle$ from Learner 1.}
\item{Learner 2 decrypts the aggregate with its private key, adds its model parameter to the aggregate and re-encrypts it with the public key of Learner 3.}
\item{Learner 2 calls post\_aggregate(2, 3, $enc_3\langle agg_2\rangle$), and so on, until we get to Learner $n$.}
\item{Learner $n$ calls get\_aggregate($n$) and receives \\
$enc_n\langle agg_{n-1}\rangle$ from Learner $n-1$.}
\item{Learner $n$ decrypts the aggregate with its private key, adds its model parameter to the aggregate and re-encrypts it with the public key of Learner 1.}
\item{Learner $n$ calls post\_aggregate($n$, 1, $enc_1\langle agg_n\rangle$)}
\item{Learner 1 receives $enc_1\langle agg_n\rangle $ from its call to get\_aggregate(1)}
\item{Learner 1 decrypts $enc_1\langle agg_n\rangle $ with its private key, subtracts $R$ and divides by $n$.}
\item{Learner 1 calls post\_average($(agg_n-R)/n$)}
\item{All learners receive $(agg_n-R)/n$ in the call to get\_average()}
\end{enumerate} 

Looking at the messages that need to be sent and assuming the public
key exchange has already accured (which is 2 messages per node,
registration and retrieval\footnote{Public key exchange does not have to be done
for every aggregation round, just once for each set of nodes you want to incorporate in the aggregation.}), both the initiator and the
non-initiator nodes have to send 4 messages.
For the initiator those are:
\begin{enumerate} 
\item{post\_aggregate}
\item{check\_aggregate}
\item{get\_aggregate}
\item{post\_average}
\end{enumerate} 
And for the non-initiator they are:
\begin{enumerate} 
\item{get\_aggregate}
\item{post\_aggregate}
\item{check\_aggregate}
\item{get\_average}
\end{enumerate} 
Hence, an aggregation requires $4n$ messages,
where $n$ is the number of learners or nodes,
for the basic algorithm.

\subsection{Progress Failover}
If a node in the chain fails to do its part the aggregation
comes to a stop. The controller can easily detect if
the aggregation is stuck in the chain and which node
failed to respond in time. For maximum flexibility we 
provide an external progress monitor that periodically 
pings the controller to see if the aggregation got stuck.
If that is the case the progress monitor will ask the
controller to notify the last node to post an aggregate
to repost its aggregate and encrypt it for the node that is
next in the chain after the failing node.
Note that this failover could have been implemented
in the nodes themselves but the nodes don't know
as easily where in the chain the process stopped and there
could be issues of multiple nodes timing out concurrently
and then creating race conditions on which nodes to jump.
It would also complicate the case where two nodes
next to each other on the chain fail simultaneously. For instance
the node that checks whether its post got consumed may also get stuck,
and then the process as a whole gets stuck, unless an external
process kicks in. Hence we decided to start off with an external
progress monitor process.

\begin{figure}[htbp]
        \centerline{\includegraphics[scale=0.3]{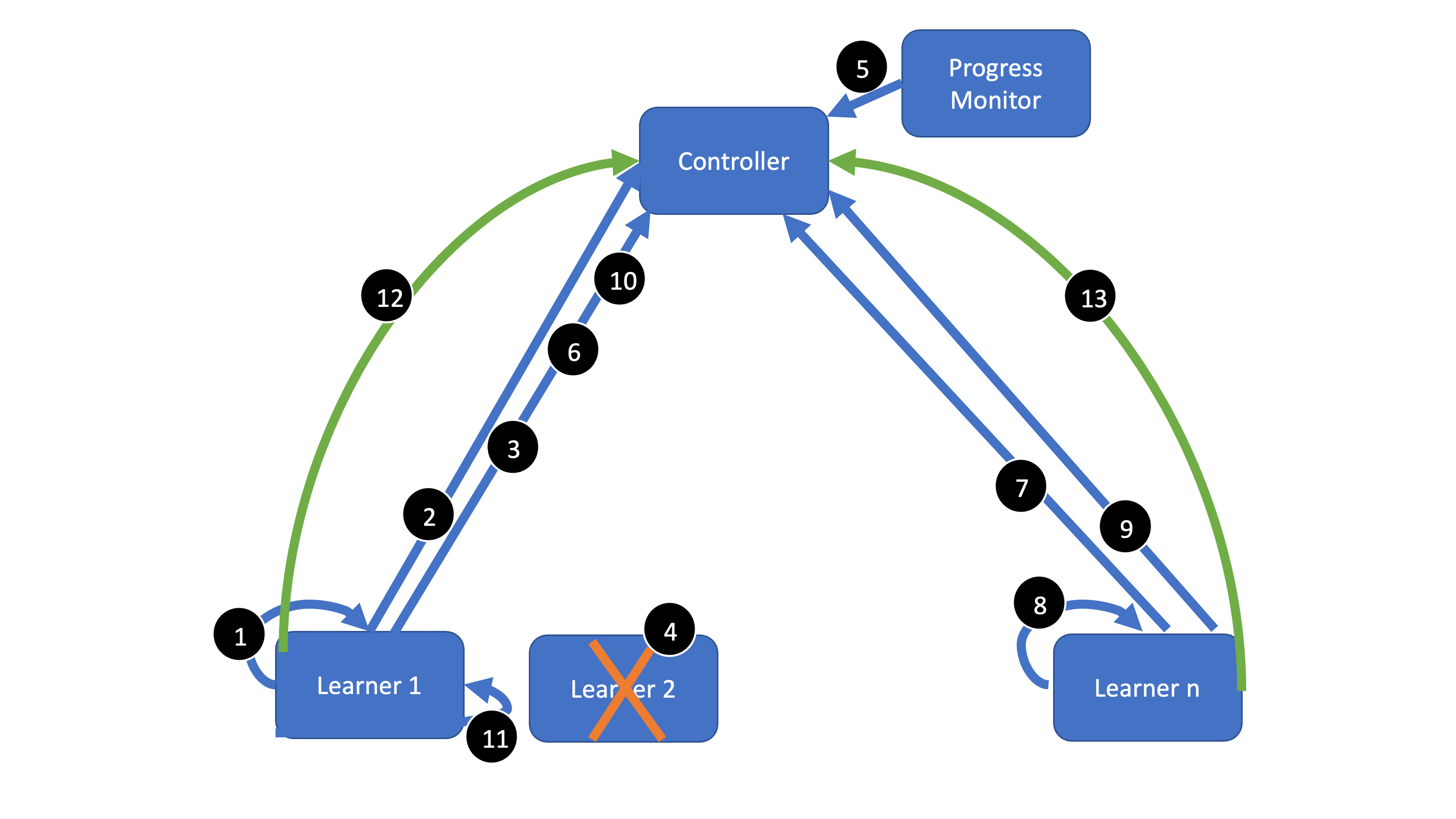}}
        \caption{Progress Failover Interactions.}
\label{failarch}
\end{figure}

The progress failover interactions can be seen
in Figure~\ref{failarch}.

\begin{enumerate} 
\item{Learner 1 generates a large random number $R$ and adds its model parameter to it, then encrypts the aggregate ($agg_1$) with the public key of Learner 2.}
\item{Learner 1 posts the encrypted aggregate 
$enc_2\langle agg_1\rangle$ to the controller by calling \\
post\_aggregate(1, 2, $enc_2\langle agg_1\rangle$)}
\item{Learner 1 starts polling for results from the final learner with check\_aggregate(2)}
\item{Learner 2 fails to do its part of the protocol}
\item{Progress Monitor detects that the protocol halted with a timeout and posts a notification to Learner 1 it needs to re-encrypt its aggregate and repost to Learner 3 as a result to the check\_aggregate call}
\item{Learner 1 calls \\
post\_aggregate(1, 3, $enc_3\langle agg_1\rangle)$}
\item{Learner $n$ calls get\_aggregate($n$) and receives \\
$enc_n\langle agg_{n-1}\rangle$ from Learner $n-1$.}
\item{Learner $n$ decrypts the aggregate with its private key, adds its model parameter to the aggregate and re-encrypts it with the public key of Learner 1.}
\item{Learner $n$ calls \\
post\_aggregate($n$, 1, $enc_1\langle agg_n\rangle$)}
\item{Learner 1 receives \\
$enc_1\langle agg_n\rangle$ from its call to get\_aggregate(1)}
\item{Learner 1 decrypts \\
$enc_1\langle agg_n\rangle$ with its private key, subtracts $R$ and divides by $n-1$.
Learner 1 is informed by the controller that only $n-1$ nodes posted aggregates}
\item{Learner 1 calls \\
post\_average($(agg_n-R)/(n-1)$)}
\item{All learners receive $(agg_n-R)/(n-1)$ in the call to get\_average()}
\end{enumerate} 

If a non-initiator node fails, two additional messages need to be sent, a new 
post\_aggregate (with the re-enrypted message retargeted next in the chain after the failed node)
and a new check\_aggregate. Hence,
the number of messages required for $n$ nodes where $f$ nodes
fail is: $4n+2f$.

In terms of drop-out rate tolerance, as long
as $n-f\geq3$ the failover will succeed. In
theory you could allow failover with 2 nodes
left too, but then both of them would
learn each other's local value by simply subtracting
their own, and hence defeating the purpose of
secure aggregation. 

\subsection{Initiator Failover}
If the initiator fails, it is not sufficient to just jump
over it in the chain, as it is the only one holding the random
secret that masks the true aggregate. Hence, in this case the
aggregation needs to be redone from the beginning.
We implement this by setting a timeout on the aggregation as a whole.
Each node would timeout after this time and then ask the controller
if they should be the new initiator. Whoever, gets assigned as the
initiator (the first one that asks the controller) starts the initiator
steps and the other nodes repeat their non-initiator steps from the
beginning.  This may mean that the failed initiator is traversed again
on the chain and you suffer from a progress failover. If this keeps
happening the nodes may have to be refreshed to permanently exclude 
the failing node from the chain.

\begin{figure}[htbp]
        \centerline{\includegraphics[scale=0.3]{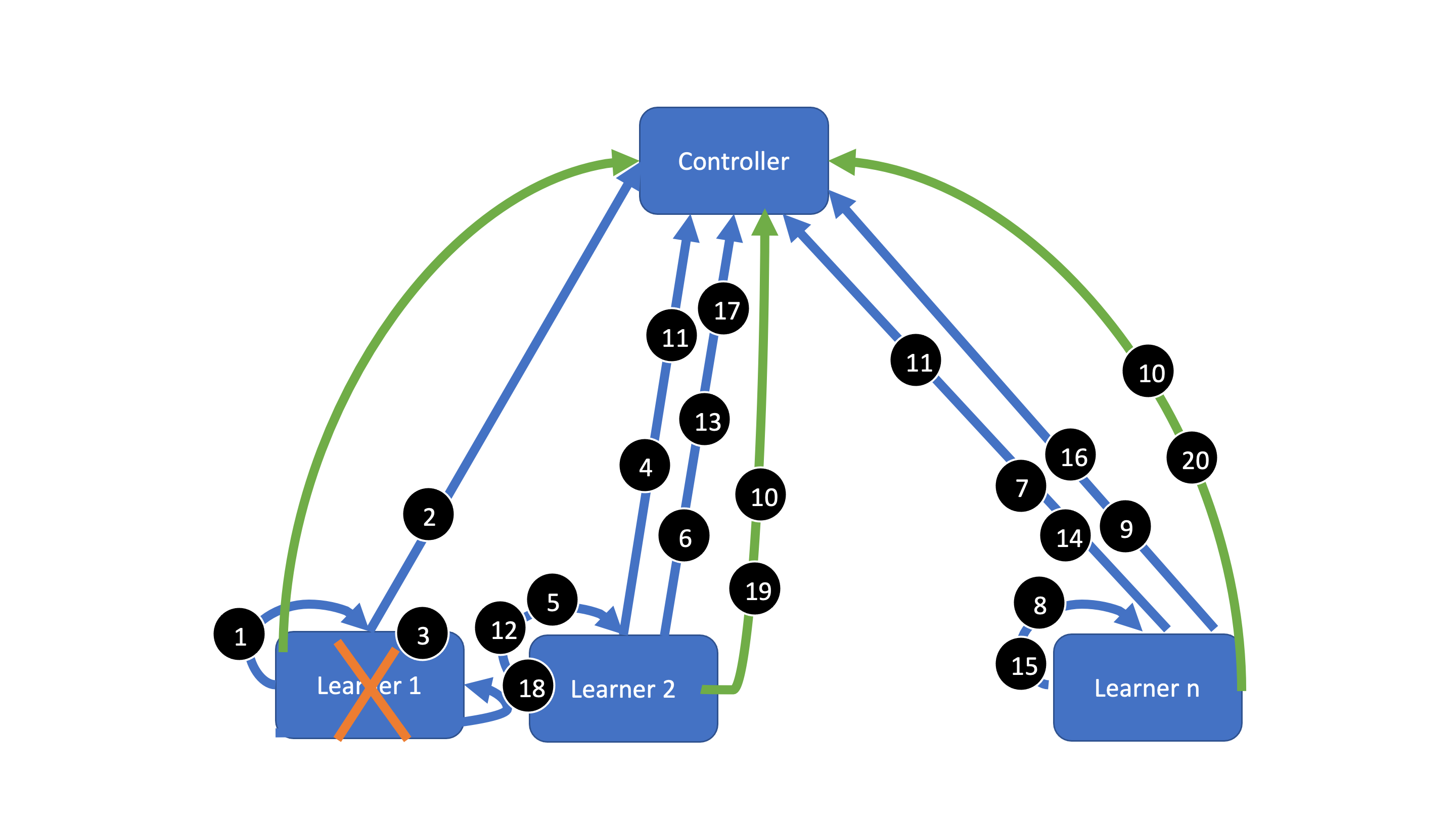}}
        \caption{Initiator Failover Interactions.}
\label{initfailarch}
\end{figure}

The initiator failover interactions can be seen
in Figure~\ref{initfailarch}.

\begin{enumerate}
\item{Learner 1 generates a large random number $R$ and adds its model parameter to it, then encrypts the aggregate ($agg_1$) with the public key of Learner 2.}
\item{Learner 1 posts the encrypted aggregate \\
$enc_2\langle agg_1\rangle$ to the controller by calling \\
post\_aggregate(1, 2, $enc_2\langle agg_1\rangle$)}
\item{Learner 1 crashes and is unable to fulfil its protocol responsibilities}
\item{Learner 2 calls get\_aggregate(2) and receives \\
$enc_2\langle agg_1\rangle$ from Learner 1}
\item{Learner 2 decrypts the aggregate with its private key, adds its model parameter to the aggregate and re-encrypts it with the public key of Learner 3.}
\item{Learner 2 calls \\
post\_aggregate(2, 3, $enc_3\langle agg_2\rangle$)}
\item{Learner $n$ calls get\_aggregate($n$) and receives \\
$enc_n\langle agg_{n-1}\rangle$ from Learner $n-1$.}
\item{Learner $n$ decrypts the aggregate with its private key, adds its model parameter to the aggregate and re-encrypts it with the public key of Learner 1.}
\item{Learner $n$ calls \\
post\_aggregate($n$, 1, $enc_1\langle agg_n\rangle$)}
\item{All learners time out in their call to get\_average}
\item{All learners call should\_initiate() and only one gets back yes, in this case Learner 2}
\item{Learner 2 generates a large random number $R’$ and adds its model parameter to it, then encrypts the aggregate ($agg_2$) with the public key of Learner 3.}
\item{Learner 2 posts the encrypted aggregate \\
$enc_3\langle agg_2\rangle$ to the controller by calling \\
post\_aggregate(2, 3, $enc_3\langle agg_2\rangle$)}
\item{Learner $n$ calls get\_aggregate($n$) and receives \\
$enc_n\langle agg_{n-1}\rangle$ from learner $n-1$.}
\item{Learner $n$ decrypts the aggregate with its private key, adds its model parameter to the aggregate and re-encrypts it with the public key of Learner 2.}
\item{Learner $n$ calls \\
post\_aggregate($n$, 2, $enc_2\langle agg_n\rangle$)}
\item{Learner 2 receives \\
$enc_2\langle agg_n\rangle$ from its call to get\_aggregate(2)}
\item{Learner 2 decrypts \\
$enc_2\langle agg_n\rangle$ with its private key, subtracts $R’$ and divides by $n-1$}
\item{Learner 2 is informed by the controller that only $n-1$ nodes posted aggregates}
\item{Learner 2 calls \\
post\_average($(agg_n-R’)/(n-1)$)}
\item{All learners receive $(agg_n-R’)/(n-1)$ in the call to get\_average()}
\end{enumerate} 

If the initiator fails all nodes need to send an additional
message (should\_initiate), and then start over.
Hence with $i$ initiator failures, $n$ nodes and $f$ progress
failures (per round) the number of messages that need to be sent is:
$(i+1)(4n+2f+in)$

\subsection{Subgrouping}
To parallelize the aggregation we allow the nodes to be subdivided
into subgroups with an initiator per subgroup. Aggregation for 
individual subgroups
may then proceed in parallel. The average
calculation will wait for all initiators to post their updates
and then an average of the posted averages will be returned
to clients from all groups. Note that we can give the same
privacy guarantees with this setup as long as each subgroup
has at least three members. This parallelization may also
be used for reliability where a single node failure
does not break the entire aggregation, just a single
subgroup.
In terms of messages being sent only a single
additional message per
group is needed, as the initiators only have the group
average they need to call get\_average after they call
post\_average in order to get the global average.
Hence with $g$ subgroups the number of messages required
is:
$(i+1)(4n+2f+in+g)$

\subsection{Weighted Averaging}
We support secure aggregation of feature vectors. It is common
that local learners create averages themselves from local models, which
in turn may be based on varying cardinality. E.g. if one learner
shares an average based on 1000 values and another shares an average
based on 10000 values the average computed but the secure aggregation
algorithm will differ from the true average.
We address this by allowing weighted averaging, where each node
can submit not just its feature vector to the secure aggregation but also
its weight, denoting the number of samples its aggregate was computed
from. Instead of the average the aggregate should be submitted with the weight.
Now as the final average the nodes will get the average of all the aggregates
as well as the average of all the weights and can then simply divide
the former with the latter to get the true average without having to reveal
how many samples each node contributed. 
Note, using weighted averaging does not require any additional
messages to be sent, but required an additional feature to be used
in the aggregate feature vector.

\subsection{Symmetric-Key Encryption}
PKI encryption with public-private keys has limitation in terms of
encryption and decryption speed as well as the payload size that 
can be encrypted. Larger payloads require larger keys which makes
the encryption and decryption slower. To mitigate this problem
we can submit a randomly generated symmetric key encrypted with the public key
of the next node on the chain and then encrypt the feature 
vector with the symmetric key.
The next node in the chain then first needs to decrypt the symmetric
key with its private key and then use the decrypted symmetric key
to decrypt the feature vector. This allows for much larger
feature vectors to be encrypted much faster.

\subsection{Symmetric-Key Pre-Negotiation}
On some constrained devices it may be too time consuming to even decrypt
the randomly generated symmetric key in each aggregation step.
In that case the symmetric key exchange may happen out of band.
Each node will generate $n$ random symmetric keys, and then
encrypt key $i={1..n}$ with the public key of node $n$, and post
all encrypted keys to the controller. 
Then when all nodes have posted their encrypted keys, all nodes pulls
down the encrypted key posted by the node next on the chain for its node
and decrypts it with its private key and caches it for later use during
the aggregation. During the aggregation instead of encrypting the feature
vector with the public key for the node next on the chain the payload
can be encrypted with the symmetric key that was cached. The receiving node
then checks where the aggregate came from and then picks the symmetric key
it generated for that node, which it stored locally to decrypt the feature
vector received.

\subsection{Pubsub Design}
\label{sec:pubsub}
Our controller supports long polling by default which means that
all nodes will open a connection to the controller and wait for a result
to come in. Hence connection establishment will not be part of the critical
path of the chain aggregation. However it could result in a lot of concurrent
connections to the controller, which ultimately would exhaust the connection
resources. We propose to mitigations to this effect.
Since the nodes at the end of the chain only need to engage at the very end of
the aggregation they can hold off on polling the controller until it is their turn.
By estimating the overall aggregation time for a cluster of nodes the nodes can hold
off based on which cluster they are assigned to before they start polling the server.
This essentially allows us to stagger the engagement with nodes. If course if the progress
estimate is off the resources may still be exhausted or the process takes longer than
needed. Staggering bigger clusters of nodes helps but may ultimately also fail. 
Hence, we also propose to separate out the notifications from the controller, and
the nodes will not poll the controller directly but wait for notifications from
on external notification system when the controller has some data for a node before
they poll the controller.

\subsection{Hierarchical Federation}
Another approach to scalability is to have many controllers that are federated
so that child controllers post their aggregate to the parent controller. This posting
does not have to be encrypted as it is already anonymized over learners, but it needs
to be coordinated. We hence also allow coordinated posting and retrieval of aggregates
across controllers, similar to the last step of the secure aggregation where an average
is posted.

\section{Edge Compute Learner Evaluation}\label{sec:simulations}
In this section we present performance results from
using a Python client implementation of our algorithm as well
as benchmarks and a collocated Python server on an 8 quadcore CPU 1-1.4Ghz 
Linux Ubuntu 20 desktop PC, where
each learner node is run concurrently in separate threads
in the same experiment process. Due to the single-machine nature
of the setup
we are limited to running about 100 concurrent nodes before
we run out of capacity, and the subgroup feature does not
add much value as everything runs on the same machine. We
hence defer the subgroup evaluation to the experiment section
below. 
Both the clients and server run in docker containers to allow for easy resetting of state between benchmark runs.
We compare our approach with (SAFE) and without (SAF) encryption to a
benchmark approach that simply posts parameters to a central controller and 
retrieves averages (INSEC). We have also invented the Practical
Secure Aggregation algorithm proposed by google (BON). Since the BON
benchmark does not scale so well in number of nodes, we perform scalability tests
both with and without it. We are interested in how the approach scales
in terms of nodes as well as features. Each condition runs in 30 repeats,
and given that the variance is so low that 99\% error bands with a z-distribution 
assumption are not visible, we display bands for $3\sigma$. Note, the error bars are hence more of a 
way to see trends in variance as opposed to statistical significance, in this case.

\subsection{Node Scalability}
\begin{figure}[htbp]
\centering
    \begin{minipage}{0.45\textwidth}
        \centering
        \includegraphics[scale=0.5]{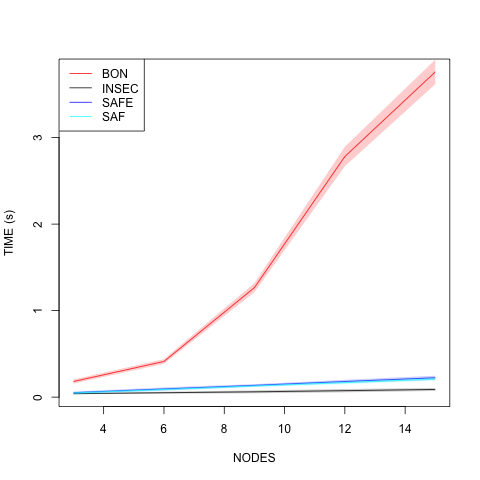}
        \caption{Edge. BON 1 Feature.}
		\label{nodesminfeatures}
	\end{minipage}\hfill
    \begin{minipage}{0.45\textwidth}
        \centering        
        \includegraphics[scale=0.5]{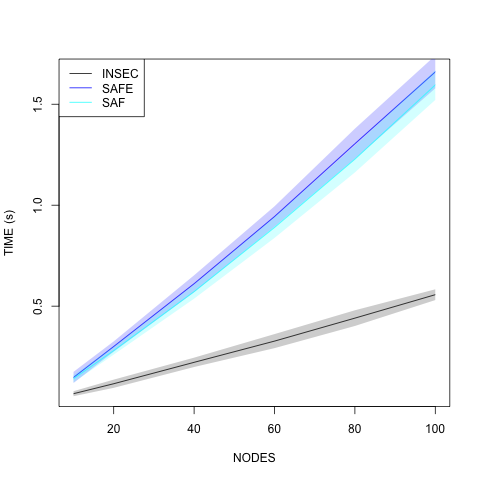}
        \caption{Edge. 1 Feature.}
		\label{nodesscaleminfeatures}
	\end{minipage}
\end{figure}

In Figure~\ref{nodesminfeatures} we can see that BON starts deteriorating in performance
already at 8-10 nodes and shows an overhead of close to 40x compared to INSEC for
15 nodes (and a single feature). For SAFE the equivalent overhead is just under 3x.
We note that we are not claiming that BON can scale beyond these numbers, but that
SAFE is more resource efficient in a small set up. Furthermore, the graph shows that
the overhead of encryption is also negligible (SAF vs SAFE) in the case of a small
feature set.

Given that BON does not scale well in this setting we wanted to 
test the limits of the SAFE
method in a larger deployment with up to 100 nodes. In \ref{nodesscaleminfeatures}
we note that the linear increase continues and the overhead between INSEC and SAFE
is still around 3x for 100 nodes (and 1 feature).

\begin{figure}[htbp]
\centering
    \begin{minipage}{0.45\textwidth}
        \centering
        \includegraphics[scale=0.5]{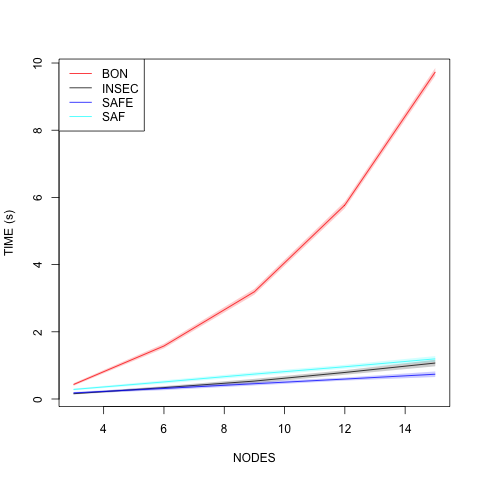}
		\caption{Edge. BON 10000 Features.}
		\label{nodesmaxfeatures}	
	\end{minipage}\hfill
    \begin{minipage}{0.45\textwidth}
        \centering        
        \includegraphics[scale=0.5]{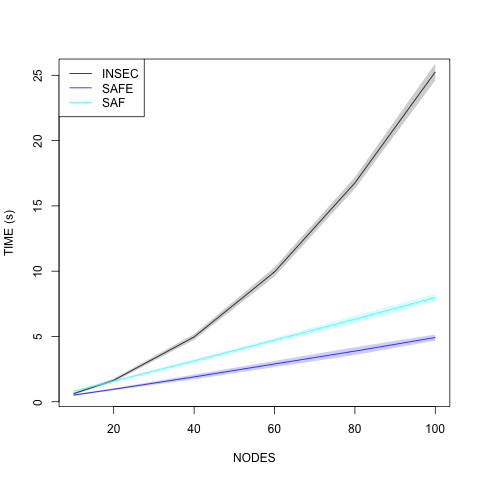}
        \caption{Edge. 10000 Features.}
		\label{nodesscalemaxfeatures}
	\end{minipage}
\end{figure}

If we increase the number of features to 10000 we can see in \ref{nodesmaxfeatures} 
and \ref{nodesscalemaxfeatures} that the trends stay the same. The BON overhead is
about 13x in the 10000 features 15 node case compared to an improvement
of about 30\% for SAFE over INSEC. This shows that SAFE encryption improved
scalability due to compression for large feature vectors. This becomes even
starker in the 100 node 10000 feature case where SAFE outperforms INSEC by a factor
of 5x. 

\subsection{Feature Scalability}
\begin{figure}[htbp]
\centering
    \begin{minipage}{0.45\textwidth}
        \centering
        \includegraphics[scale=0.5]{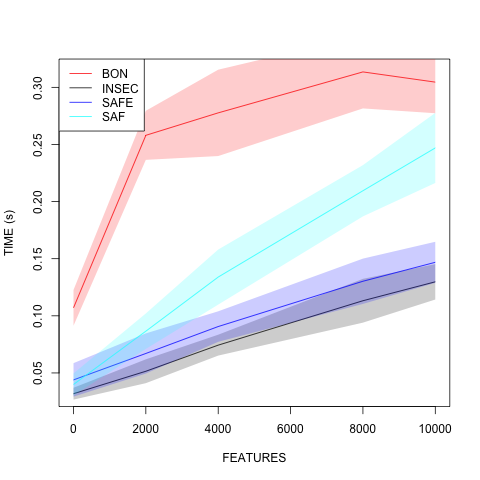}
		\caption{Edge. BON 3 Nodes.}
		\label{featuresminnodes}	
	\end{minipage}\hfill
    \begin{minipage}{0.45\textwidth}
        \centering        
        \includegraphics[scale=0.5]{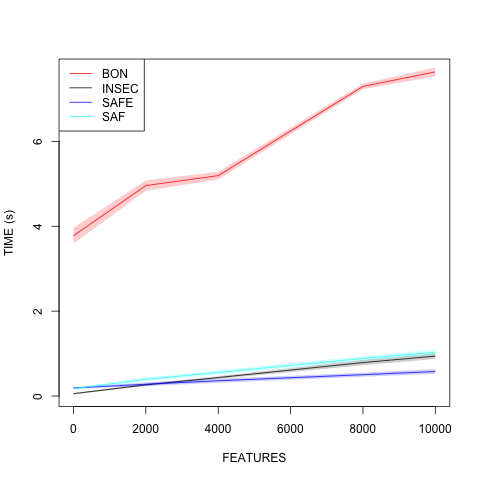}
        \caption{Edge. BON 15 Nodes.}
		\label{featuresmaxnodes}
	\end{minipage}
\end{figure}

\begin{figure}[htbp]
        \centerline{\includegraphics[scale=0.5]{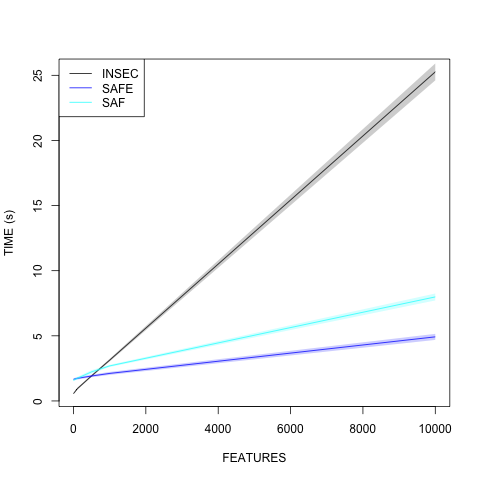}}
        \caption{Edge. 100 Nodes.}
\label{featuresscalemaxnodes}
\end{figure}

We now study how the algorithms scale in terms of feature vector sizes in
Figures~\ref{featuresminnodes},~\ref{featuresmaxnodes}, and~\ref{featuresscalemaxnodes}. We can see for 10000 features and 3 nodes INSEC is still faster than SAFE,
but (a) with 15 nodes, the crossover is somewhere around 2000 features, and (b) with 
100 nodes, the crossover occurs around 100 features.

To summarize, we can see that the encryption mechanism in SAFE helps with feature
scalability as it also compresses the payload. The BON mechanism scales
much worse in number of nodes given that all nodes need to submit both their
weights as well as their secret masks to the controller and the controller
also needs to be involved in the aggregation, as opposed to just passing along
messages in our method (SAFE).

We note that to support encryption of larger payloads efficiently, 
i.e. bigger feature vectors, we use
RSA encryption of a symmetric AES key that then is used to decrypt the encrypted
feature vector. Both the encrypted message key and the encrypted feature
vectors are exchanged between the nodes in each aggregation step.
The aggregation payload is opaque to the controller and any JSON object may be passed along. 
The only assumption is that all the nodes participating in the aggregation
need to know how to decrypt and encrypt as well as parse the payload.

\subsection{Failover Overhead}
With SAFE, if a node fails the controller needs to direct the next node in the chain
to pick up the aggregate, and let the initiator know how many
nodes successfully posted aggregates in order to complete the
calculation of the average. With BON, if a node fails all remaining nodes
need to report a separate secret to be added to the global aggregate
based on which node failed. Next, we measure the overhead that a failing
node imposes on the calculation of the global average. Note, if the initiator
fails in the SAFE method the whole aggregation just needs to restart, so the
failure completion time is 2x, if x is the time to compute the average without
failure. Here, we focus on a non-initiator failure, which is the more
interesting case.

A failure is detected by setting a timeout on getting the result. Clearly, the
best timeout to set depends on the number of nodes and features in the aggregation,
i.e. the expected aggregation time. We measured the expected completion time
without failure, and then made a prediction using a second degree polynomial
fit and added 4 seconds of safety margin time, to avoid false positives.
Here we only use a single feature, as the number of features does not impact
the failover scenario and it allows us to run aggregations faster and with more nodes.
To measure the failover overhead we subtract the expected failure timeout time (when nodes just wait for failed nodes to complete)
from the overall aggregation time. For BON
this is a global wait time and for SAFE this is the progress timeout per failed node. To facilitate
apples-to-apples comparisons we kept the sum of all failed node timeouts in SAFE the same as the global BON timeout.

To simulate a failure, we complete the public key 
exchange step for all nodes before taking out
nodes 4 to 6 in the chain 
(any non-initiator nodes could have been picked)
and starting the aggregation process. We then compare
the aggregation time for a given number of completed nodes, e.g.,
the aggregation time for 21 nodes without failures is compared to
the aggregation time for 24 nodes with 3 failures~\footnote{Without this normalization, failure scenarios tend to scale better as fewer nodes are involved.}.

\begin{figure}[htbp]
\centering
    \begin{minipage}{0.45\textwidth}
        \centering
        \includegraphics[scale=0.5]{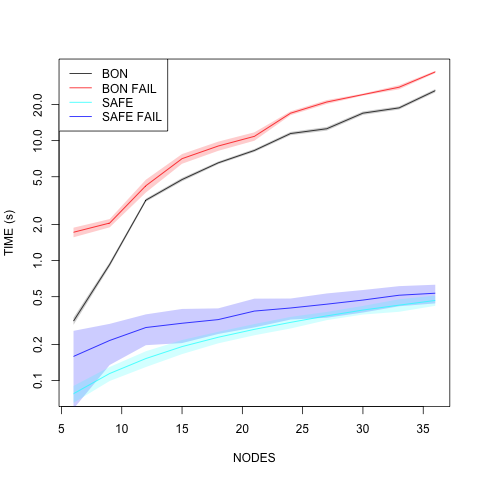}
		\caption{Edge. Failover.}
		\label{failover}	
	\end{minipage}\hfill
    \begin{minipage}{0.45\textwidth}
        \centering        
        \includegraphics[scale=0.5]{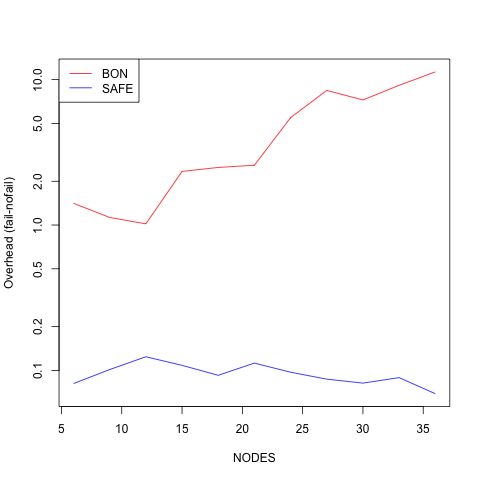}
        \caption{Edge. Failover Overhead.}
		\label{overhead}
	\end{minipage}
\end{figure}

Figures~\ref{failover} and~\ref{overhead} (error bands are $3\sigma$, and y-axis in log-scale) show how
BON performance both with and without failover scales super linearly
with nodes, whereas both SAFE and SAFE with failover scales linearly. 

With 24 nodes the BON failover to SAFE failover
aggregation time ratio exceeds that of BON without failover
to SAFE without failover (42x vs 38x).
At 36 nodes the corresponding ratios are 70x and 56x.

For comparison, we also note here that the authors of~\cite{so2021},
showed a 40x improvement with 200 nodes, i.e. an order of
magnitude more nodes were needed to reach the same level of improvements.

\section{Deep-Edge Constrained Device Learner Evaluation}\label{sec:experiments}
Next we explore the performance of our method on
an embedded platform, OpenWrt. We implemented 
the aggregation client using busybox coreutils, curl and
openssl and deployed it on twelve Archer C7 TP-Link
Wi-Fi routers. We use the symmetric key pre-negotiation
feature and exchange symmetric keys before the
aggregation as RSA key decryption is very slow
on these devices. The 12 routers are connected
to a LAN over an Ethernet backhaul. The controller
is deployed on the same LAN on a MacBook Pro PC,
which also coordinates the aggregation experiments
over ssh. Generating random numbers is also quite
slow on this platform so only a single seed is used
regardless of the number of features aggregated in
the initiator. Given the 12 learners, we explore
groupings of $1 \times 12$, $2 \times 6$, $3 \times 4$,
and $4 \times 3$. The BON algorithm client was 
not implemented on this platform but both SAF and INSEC 
were ported as well.
Each condition runs in 5 repeats, and given that the variance is so 
low that 99\% error bands with a t-distribution 
assumption are not visible, we display bands for $4\sigma$.

\subsection{Node Scalability}
\begin{figure}[htbp]
\centering
    \begin{minipage}{0.45\textwidth}
        \centering
        \includegraphics[scale=0.5]{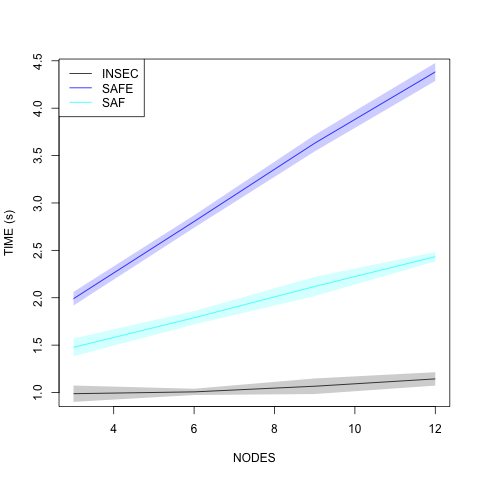}
		\caption{Deep-Edge. 1 Feature.}
		\label{expnodesminfeatures}
	\end{minipage}\hfill
    \begin{minipage}{0.45\textwidth}
        \centering        
        \includegraphics[scale=0.5]{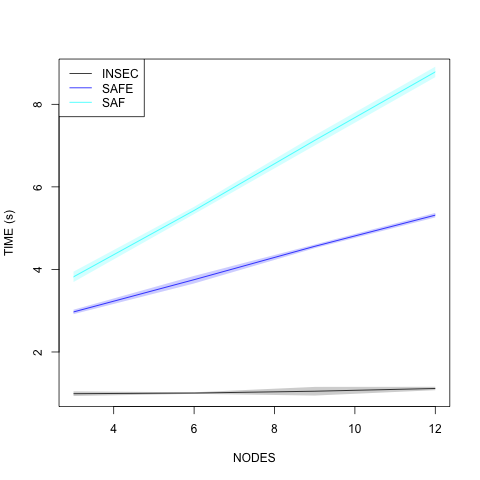}
		\caption{Deep-Edge. 20 Features.}
		\label{expnodesmaxfeatures}	
	\end{minipage}
\end{figure}

First, looking at node scalability, we see in
Figure~\ref{expnodesminfeatures} that SAFE has an
overhead of about 2x with 3 nodes and 4.5x with 12 nodes,
compares to INSEC (with a single feature). Although
SAF suffers with 20 features, the SAF vs INSEC
differences are roughly the same with 20 features in
Figure~\ref{expnodesmaxfeatures}.

\subsection{Feature Scalability}
\begin{figure}[htbp]
\centering
    \begin{minipage}{0.45\textwidth}
        \centering
        \includegraphics[scale=0.5]{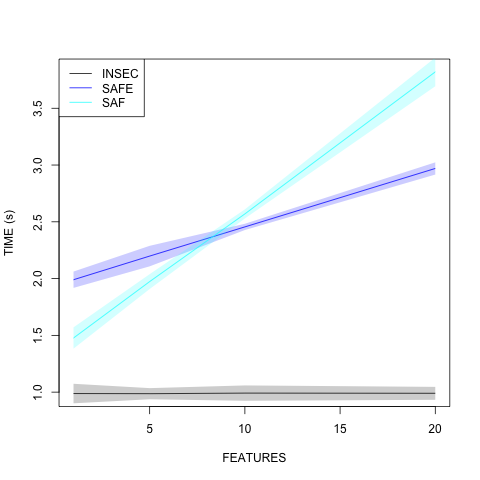}
        \caption{Deep-Edge. 3 Nodes.}
        \label{expfeaturesminnodes}
	\end{minipage}\hfill
    \begin{minipage}{0.45\textwidth}
        \centering        
        \includegraphics[scale=0.5]{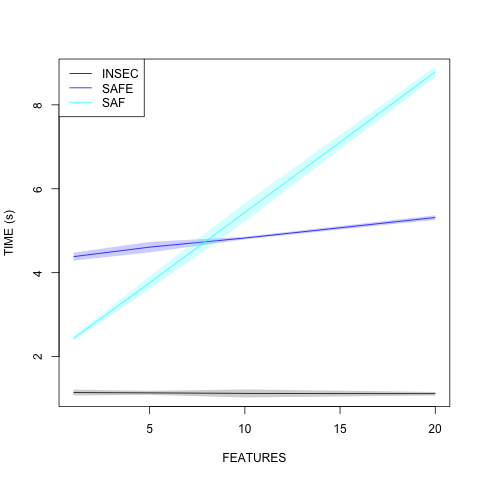}
        \caption{Deep-Edge. 12 Nodes.}
		\label{expfeaturesmaxnodes}
	\end{minipage}
\end{figure}

In Figures~\ref{expfeaturesminnodes} and~\ref{expfeaturesmaxnodes}
we can see the crossover between SAF and SAFE 
performance is between 5 and 10 features both for 3 and 12 nodes.

\subsection{Subgrouping Improvement}
\begin{figure}[htbp]
\centering
    \begin{minipage}{0.45\textwidth}
        \centering
        \includegraphics[scale=0.5]{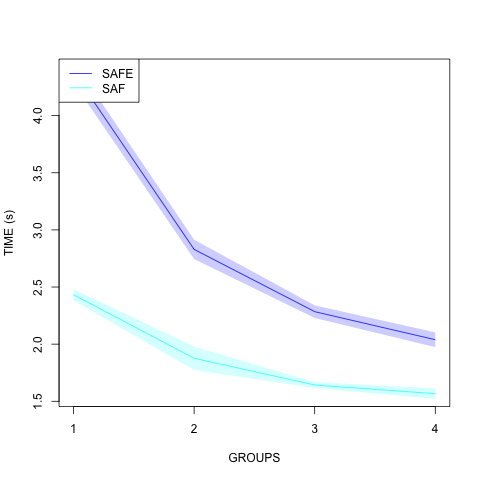}
        \caption{Deep-Edge. 12 Nodes 1 Feature.}
		\label{groupsminfeatures}
	\end{minipage}\hfill
    \begin{minipage}{0.45\textwidth}
        \centering        
        \includegraphics[scale=0.5]{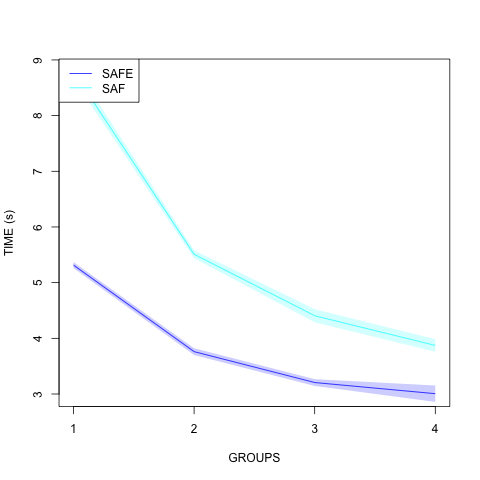}
        \caption{Deep-Edge. 12 Nodes 20 Features.}
		\label{groupsmaxfeatures}
	\end{minipage}
\end{figure}

How can we improve the performance for large number of nodes
with SAFE? We can introduce sub groups that aggregate
in parallel as alluded to in Section~\ref{sec:design}. 
Figure~\ref{groupsminfeatures} shows we can improve the
aggregation time from about 4.5 seconds down to about 2 seconds
with four parallel group vs a single group with SAFE and 1 feature.
For 20 features (See Figure~\ref{groupsmaxfeatures}), we can get it
down from 5.5 to 3 seconds with four groups.

From these results we can see again that encryption helps with compression
and scalability and that subgrouping is efficient in improving the
parallelism and performance of aggregations, as it can more than double the
aggregation throughput.

\section{Discussion and Conclusion}\label{sec:discussion}
We have shown the practical benefits of a secure aggregation
mechanism that relies on a single mask by an initiator in
a public-key-encryption-secured chain of participating
learners. Not only is it faster in the most common case
of all learners contributing successfully to the protocol,
but it is also able to recover faster if some node fails
to contribute its value.

We show that the algorithm, with some tweaks to pre-negotiate
symmetric keys, can be implemented efficiently on constrained
devices such as OpenWrt Wi-Fi access points, by simply relying
on coreutils primitives and OpenSSL.

In deployments with a large number of nodes we also show that
subgrouping can improve performance. This method has its limits
though as the single coordinator may at some point still
become a bottleneck. In such scenarios a hierarchical federated
learning topology may be deployed to completely detach
the running of the protocol in the subgroups from each other.

We should also note that if a large number of nodes fail to
complete their protocol step, the aggregation may not be as efficient
as each failing node would cause a hiccup in the progress
independently along the chain. 
This effect could be mitigated by having a way of checking the health
of nodes and remove them from the chain pro-actively, and periodically
refresh the chain to remove nodes that are contributing too intermittently.

The performance of our approach relies on an efficient notification or
pubsub system. If the number of nodes outgrow what the pubsub system can
handle efficiently the performance suffers significantly. This could
be addressed by having each node predict when it is its turn to
contribute an aggregate so that not all nodes in the chain overwhelm
the controller or pubsub system all at once. At some point when
you have a very large pool of nodes to compute an average over,
it is probably more practical to deploy some noise mechanism, such
as differential privacy. Based on our experiments, the SAFE method
could be deployed efficiently and with good privacy guarantees
when you have subgroups of sizes between 3-100 nodes. Of course
if you deploy hierarchical federation that could still 
cover very large pools, e.g. with thousands of nodes. However,
it would probably not be the right approach when you have millions
of nodes.

We note that a node that colludes with another node on the chain
can infer the average of the intermediate nodes on the chain.
So if node $n$ colludes with node $n-2$ they will be able to
infer the local value of node $n-1$. Protection against this kind
of collusion would require additional use of masking which
is outside the scope of our current work. In general,
we assume honest but curious behavior and no collusion
between nodes. You could randomize the order between each round to
limit the likelihood of two colluding nodes being able to get useful
data from intermediaries on a consistent basis. Recall that
secure aggregation is typically deployed as part of a gradient descent
iteration, so only being able to infer values from a subset of a large
number of iterations would still offer some protection of privacy of 
the local data, assuming a sufficiently large number of nodes participate
in the aggregation.

Finally, there is the issue of the distribution and security of private keys. While they offer the above-mentioned improvements over PKI’s, they are still subject to the usual attacks that are thoroughly described in the literature. One way to improve their security is using quantum internet protocols that can distribute keys through quantum channels that are provable-and not just algorithmically secure. An example of such technology using photons and operating at the level of the transport layer of the internet has been shown in~\cite{huberman2020}.

\bibliographystyle{IEEEtran}
\bibliography{related}
\newpage
\appendix
\section{Controller Python Flask Implementation}\label{controllerimplementation}
\lstset{language=Python,basicstyle=\scriptsize\ttfamily,showspaces=false,showstringspaces=false} 
\begin{lstlisting}[backgroundcolor = \color{lightgray}, framexleftmargin = 0.2em, framexrightmargin = 1em, multicols=2]
@app.route('/should_initiate',methods=['POST'])
def should_initiate():
    with lock:
      data = request.get_json(force=True)
      node = data["node"]
      current_time = time.time()
      group = 1
      if "group" in data:
        group = data["group"]
      if not group in average: 
        init_average(group, node)
        return json.dumps({"init": True})
      if (current_time - average[group]["time"]) \
               > config["aggregation_timeout"]:
        init_average(group, node)
        return json.dumps({"init": True})
      return json.dumps({"init": False})

@app.route('/post_aggregate',methods=['POST'])
def post_aggregate():
    data = request.get_json(force=True)
    with lock:
      from_node = data["from_node"]
      to_node = data["to_node"]
      group = 1
      if "group" in data:
        group = data["group"]
      if group in average and \
         average[group]["initiator"] == from_node:
        init_average(group, from_node)
      elif group not in average:
        init_average(group, from_node)
      if not group in aggregate:
        aggregate[group] = {}
      aggregate[group][to_node] = \
         {"aggregate": data["aggregate"],
          "time": time.time(), 
          "from_node": from_node}
      group_stats[group]["posted"] += 1
      if group not in repost_aggregate:
        repost_aggregate[group] = {}
      repost_aggregate[group][from_node] =  \
          {"status": "consumed"}
      repost_aggregate[group][to_node] =  \
          {"status": "empty"}
      return json.dumps(data)

def internal_check_aggregate(data):
    group = 1
    if "group" in data:
      group = data["group"]
    node = data["node"]
    result = {"status": "empty"}
    if group in repost_aggregate and \
          node in repost_aggregate[group]:
      result = repost_aggregate[group][node] 
      del repost_aggregate[group][node]
    return result

def poll_internal(data, func):
    TIMEOUT = config["poll_time"]
    WAIT_TIME = config["yield_time"]
    empty = True
    start_time =  time.time()
    with lock:
      result = func(data)
    empty =  ("status" in result) and \
          (result["status"] == "empty")
    while empty and \
         (time.time() - start_time) < TIMEOUT:
      time.sleep(WAIT_TIME)
      with lock:
        result = func(data)
        empty =  ("status" in result) and \
            (result["status"] == "empty")
    return result

@app.route('/check_aggregate',methods=['POST'])
def check_aggregate():
    data = request.get_json(force=True)
    result = poll_internal(data, 
                      internal_check_aggregate)
    return json.dumps(result)

def internal_get_aggregate(data):
    result = {"status": "empty"}
    group = 1
    if "group" in data:
      group = data["group"]
    if group in aggregate and \
          data["node"] in aggregate[group]:
      result = {"status": "ok"}
      if "aggregate" in \
           aggregate[group][data["node"]]:
        result["aggregate"] = \
       aggregate[group][data["node"]]["aggregate"]
      if "from_node" in \
           aggregate[group][data["node"]]:
        result["from_node"] = \
       aggregate[group][data["node"]]["from_node"]
      del aggregate[group][data["node"]]
      result["posted"] = \
         group_stats[group]["posted"] - \
         group_stats[group]["skipped"] 
    return result

@app.route('/get_aggregate',methods=['POST'])
def get_aggregate():
    data = request.get_json(force=True)
    result = poll_internal(data, 
                           internal_get_aggregate)
    return json.dumps(result)

@app.route('/post_average',methods=['POST'])
def post_average():
    data = request.get_json(force=True)
    with lock:
      group = 1
      if "group" in data:
        group = data["group"]
      average[group]["average"] = data["average"]
      average[group]["status"] = "posted"
      if group not in repost_aggregate:
        repost_aggregate[group] = {}
      if "node" in data:
        repost_aggregate[group][data["node"]] = \
            {"status": "consumed"}
      return json.dumps(data)
\end{lstlisting}

%\newpage
\section{Aggregator Bash Implementation}\label{bashimplementation}
\lstset{language=bash,basicstyle=\scriptsize\ttfamily,showspaces=false,showstringspaces=false} 
\begin{lstlisting}[backgroundcolor = \color{lightgray}, framexleftmargin = 0.2em, framexrightmargin = 1em, multicols=2]
if [ ${IS_INITIATOR} -eq 1 ]; then
  R=$(${SH} ${SCRIPT_DIR}/get_random ${FEATURES} 1)
  RAW_AGG=$(vector_add "$R" "$VAL")
  if [ ${SHOULD_ENCRYPT} -eq 1 ]; then
    AGG=`encrypt ${RAW_AGG}`
  else
    AGG="${RAW_AGG}"
  fi
  curl -s -d '{"from_node":'${NODE}',"to_node": \
     '${NEXT_NODE}',"aggregate":"'"${AGG}"'", \
     "group":'${GROUPID}'}' \
     ${CONTROLLER}/post_aggregate >/dev/null
  wait_for_repost "${RAW_AGG}" ${NEXT_NODE} 
  JSON=$(wait_for get_aggregate \
   '{"node":'${NODE}',"group":'${GROUPID}'}')
  AGGREGATE=`get_json aggregate "$JSON"`
  POSTED=`get_json posted "$JSON"`
  if [ ${SHOULD_ENCRYPT} -eq 1 ]; then
    AGGREGATE=`decrypt $AGGREGATE`
  fi
  AVG=$(vector_sub "$AGGREGATE" "$R" "$POSTED")
  AVG=`echo -e "$AVG" | tr ' ' ','`
  JSON=`curl -s -d '{"average":['${AVG}'], \
    "node":'${NODE}',"group":'${GROUPID}'}' \
     ${CONTROLLER}/post_average`
  JSON=$(wait_for get_average '{}')
  AVG=`get_json_arr "$JSON"`
else
  # non-initiator
  JSON=$(wait_for get_aggregate \
   '{"node":'${NODE}',"group":'${GROUPID}'}')
  AGGREGATE=`get_json aggregate "$JSON"`
  if [ ${SHOULD_ENCRYPT} -eq 1 ]; then
    AGGREGATE="$(decrypt $AGGREGATE)"
  fi
  RAW_AGG=$(vector_add "$AGGREGATE" "$VAL")
  if [ ${SHOULD_ENCRYPT} -eq 1 ]; then
    AGG=`encrypt ${RAW_AGG}`
  else
    AGG="${RAW_AGG}"
  fi
  curl -s -d '{"from_node":'${NODE}', \
    "to_node":'${NEXT_NODE}', \
    "aggregate":"'"${AGG}"'", \
    "group":'${GROUPID}'}' \
    ${CONTROLLER}/post_aggregate >/dev/null
  wait_for_repost "${RAW_AGG}" ${NEXT_NODE} 
  JSON=$(wait_for get_average '{}')
  AVG=`get_json_arr "$JSON"`
fi
\end{lstlisting}

\end{document}